\documentclass[final,times,twocolumn]{elsarticle}

\usepackage{lineno,hyperref,subcaption,amsmath}
\journal{arXiv}

\usepackage[
separate-uncertainty=true,
per-mode=symbol-or-fraction,
]{siunitx}
\usepackage{graphicx}

\begin{document}
	
	\begin{frontmatter}
		
		\title{Techniques for the investigation of segmented sensors using the \mbox{Two Photon Absorption -- Transient Current Technique}}
		
		\author[1,2]{S. Pape\corref{mycorrespondingauthor}}
		\author[1]{E. Currás}
		\author[3]{M. Fernández García}
		\author[1]{M. Moll}
		
		\address[1]{CERN, Switzerland}
		\address[2]{TU Dortmund University, Germany}
		\address[3]{Instituto de Física de Cantabria (CSIC-UC), Spain}
		
		\cortext[mycorrespondingauthor]{Corresponding author. E-mail address: \href{mailto:sebastian.pape@cern.ch}{sebastian.pape@cern.ch}}
		
		
		\begin{abstract}
			The Two Photon Absorption - Transient Current Technique (TPA-TCT) was used to investigate a silicon strip detector with illumination from the top.
			Measurement and analysis techniques for the TPA-TCT of segmented devices are presented and discussed using a passive strip CMOS detector as an example.
			The influence of laser beam clipping and reflection is shown and a method that allows to compensate these intensity related effects for the investigation of the electric field is introduced and successfully employed. Besides, the mirror technique is introduced, which exploits reflection at a metallised back side to enable the measurement directly below a top metallisation, while illuminating from the top.
			
		\end{abstract}
		
		
	\end{frontmatter}
	
	
	\section{Introduction}
	
	The Two Photon Absorption - Transient Current Technique (TPA-TCT) was developed in the framework of the RD50 collaboration for the characterisation of silicon detectors~\cite{ivan_rd50, rogelio_rd50}.
	It is based on the well established Single Photon Absorption - Transient Current Technique (SPA-TCT)~\cite{TCT, kramberger_edge}, where lasers with wavelength within the linear absorption regime of silicon ($<\SI{1.15}{\micro\meter}$) and a continuous absorption along the beam propagation are used.
	In contrast, the TPA-TCT uses lasers with a wavelength in the quadratic absorption regime ($>\SI{1.2}{\micro\meter}$ to $<\SI{2.3}{\micro\meter}$), wherefore, a confined excess charge carrier density is only generated around the focal point, which allows to probe the silicon bulk with a three dimensional resolution.
	Equivalent to conventional TCT methods, the charge carrier drift of the generated excess charge carriers is recorded and device quantities like the charge collection time, the collected charge, or the electric field can be investigated, however in a three dimensional manner.
	Therefore, TPA-TCT can be performed with illumination from the top, back, or edge, while still maintaining the resolution along the laser beam propagation~\cite{Wiehe_thesis}, which is for SPA based methods only possible for illumination from the edge~\cite{kramberger_edge}.
	The advantage of top illumination is that the sample alignment is easier and no sample preparation, like polishing, is needed. 
	
	Segmented devices benefit in particular from a three dimensional resolution~\cite{HV-CMOS, HV-CMOS2, GARCIA2020162865}. 
	However, they can add complexity to the measurement, because different material layers are usually involved, which can lead to reflection~\cite{Pape_2022} or laser beam clipping and could thereby influence the laser intensity, i.e. the generated charge, depending on the position inside the device.
	Within the here presented study, methods for the characterisation of segmented devices are presented using a passive p-type strip CMOS detector as an example, which was developed in a collaborative project of DESY, the University of Bonn, the University of Freiburg, and the TU Dortmund University~\cite{DIEHL2022166671, DIEHL2022167031}. It is intended to be used as a cost efficient device for large detector experiments, due to the exploitation of the commercialised CMOS process.
	
	The paper is structured in the following way: first an overview about the used setup and the device under test (DUT) is given in section~\ref{sec:Experimental setup}, then analysis methods that are especially useful to segmented sensors are derived in section~\ref{sec:Analysis methods}, which is followed by a characterisation of the DUT with TPA-TCT and an application of the prior derived methods in section~\ref{sec:Results}. Finally, the results are concluded in section~\ref{sec:Conclusion}.

	\section{Experimental setup}
	\label{sec:Experimental setup}
	
\begin{figure}
		\centering
		\includegraphics[width = 0.45\textwidth]{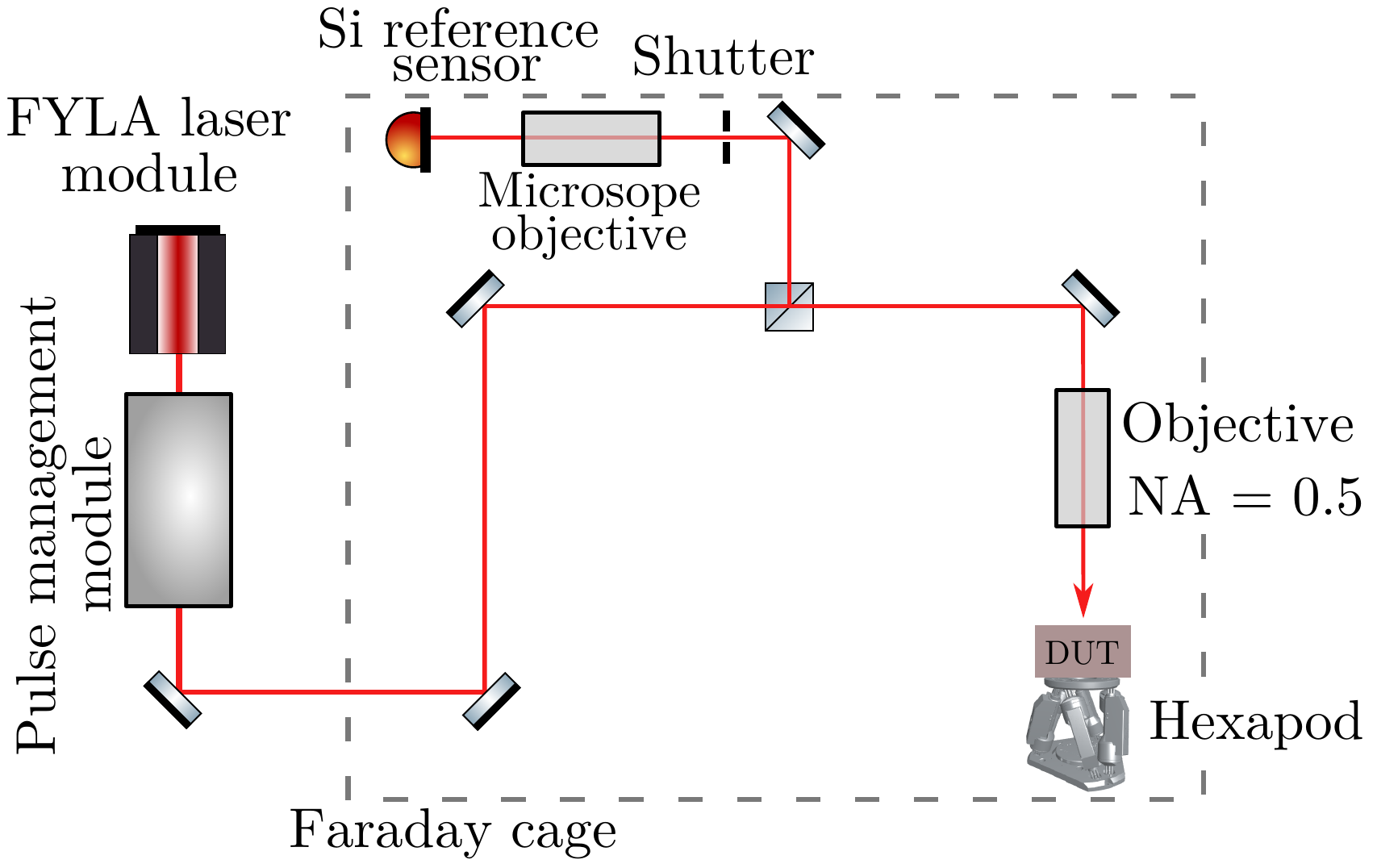}
		\caption{Schematic drawing of the used TPA-TCT setup.}	
		\label{fig:setup}
	\end{figure}

	A schematic of the used TPA-TCT setup is depicted in figure~\ref{fig:setup}.
	The setup uses the FYLA LFC1500X fibre laser module~\cite{Almagro-Ruiz:22}, which provides a $\SI{430}{\femto\second}$-pulse with a wavelength of $\SI{1.55}{\micro\meter}$, a pulse frequency of $\SI{8.2}{\mega\hertz}$, and a pulse energy of $\SI{10}{\nano\joule}$ at its output.
	Behind the output the light is coupled out from the fibre and traverses in open space towards the pulse management module.
	The pulse management module includes an acusto-optic modulator to regulate the pulse frequency and a neutral density filter to adapt the pulse energy.
	For the here presented measurements, a pulse frequency of $\SI{200}{\hertz}$ and a pulse energy of $\SI{0.22}{\nano\joule}$ (measured below the objective) is used.
	It is experimentally verified that this pulse energy is below the threshold of electron-hole plasma creation, which arise for high enough charge carrier densities~\cite{tove-seibt}.
	
	Behind the pulse management module, the light is guided inside a Faraday cage, which is used for electromagnetic shielding. 
	Inside the Faraday cage, the light passes a 50/50 beam splitter, where one arm reaches the DUT and the other arm reaches a silicon diode that is used as a reference to correct for potential energy fluctuation from the laser source.
	Highly focusing objectives are used in both arms, to increase the charge generation by TPA, as it scales quadratically with the light intensity.
	The highest possible focusing, while avoiding aberration~\cite{Wiehe_thesis, gu2000advanced}, is used to increase the TPA efficiency as much as possible.
	For the here presented measurements, an objective with a numerical aperture of 0.5 is found to be ideal.
	The objective achieves, for the alignment at hand, a beam waist of $w_0 = \SI{1.63(11)}{\micro\meter}$ and a Rayleigh length inside silicon of $z_{\mathrm{R, Si}} = \SI{13.07(105)}{\micro\meter}$ in the DUT arm. 
	The beam parameters are measured with the knife-edge technique, following the procedure described in reference~\cite{Wiehe_paper}.
	The beam parameters of the reference arm were not measured.
	
	The DUT is glued with silver epoxy to a passive readout board, which is mounted below the objective on a copper chuck.
	The chuck is thermally coupled to a Peltier element, where the hot side is cooled by a HUBER chiller.
	For the here presented measurements, the DUT is actively temperature controlled to $\SI{20}{\degreeCelsius}$, the Faraday cage is continuously flushed with dry air, and the device is illuminated from the top.
	The copper chuck is positioned on a six axes HXP50-MECA stage, which allows high precision movement and rotation along all three dimensional axes. The rotation is used to level the DUT and ensure an orthogonal incidence of the laser.
	Further information about the setup can be found in reference~\cite{Wiehe_paper}.
	
\begin{figure}
		\centering
		\includegraphics[width = 0.45\textwidth]{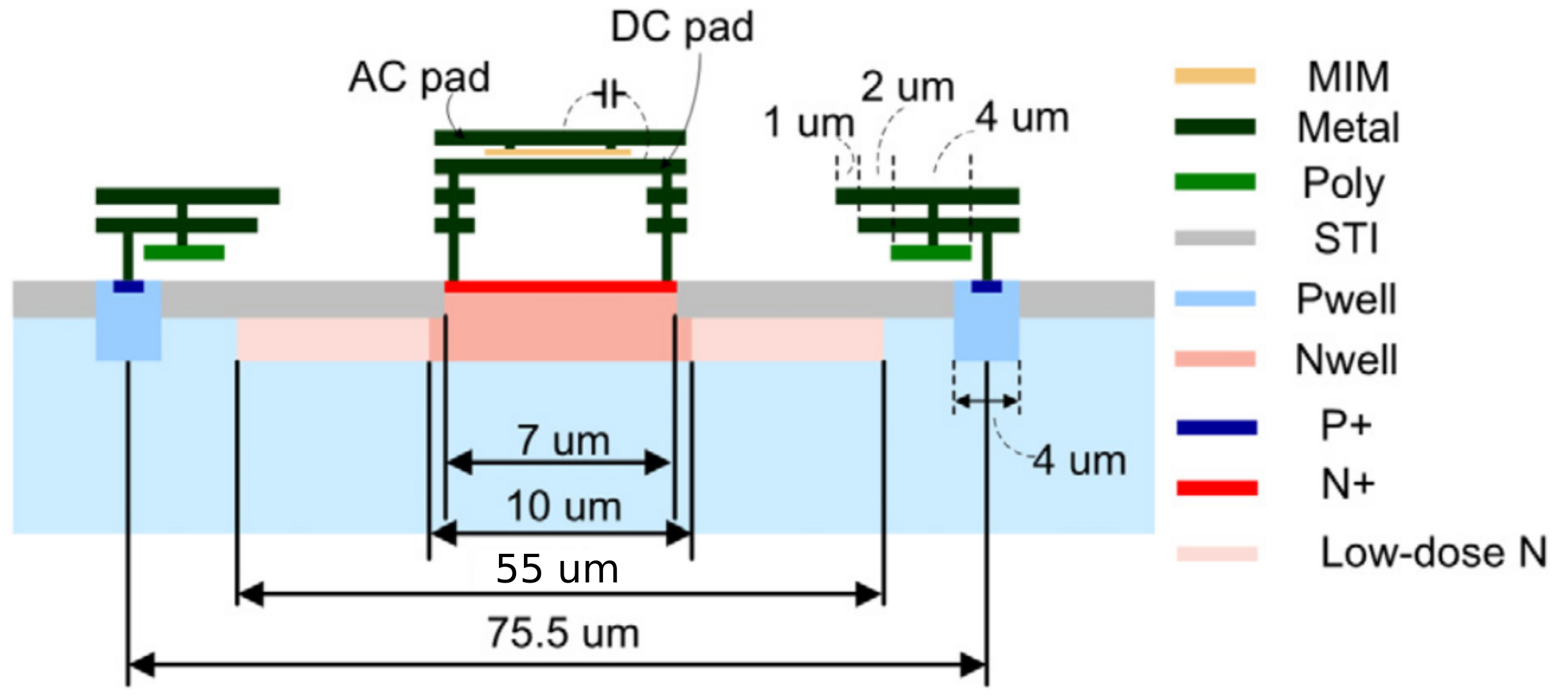}
		\caption{Schematic drawing of the implantation of the passive strip CMOS detector (taken from~\cite{DIEHL2022166671}). MIM and STI stand for metal-insulator-metal and shallow trench isolation, respectively.}	
		\label{fig:implantation}
	\end{figure}

	A passive strip CMOS detector is used, to investigate the TPA-TCT in segmented detectors.
	The device is a $\SI{150}{\micro\meter}$ thick p-type strip detector, with a pitch of $\SI{75.5}{\micro\meter}$ and $\SI{2}{\centi\meter}$ long strips. The resistivity of the wafer is \mbox{3 - $\SI{5}{\kilo\ohm\centi\meter}$} and the device is fully depleted for bias voltages above $\SI{30}{\volt}$.
	It is fabricated in a CMOS process, with $\approx\SI{1}{\centi\meter^2}$ big reticles that are stitched together.
	It has been shown that the stitching does not affect the devices performance~\cite{DIEHL2022166671, DIEHL2022167031}.
	There are three different implantation designs available: the regular, the low dose $\SI{55}{\micro\meter}$, and the low dose $\SI{30}{\micro\meter}$ design. All the here presented measurements are performed on the low dose $\SI{55}{\micro\meter}$ design, which is shown figure~\ref{fig:implantation}.
	Further information about the implantation designs can be found in the references~\cite{DIEHL2022166671, DIEHL2022167031}.
	Moreover, the back side of the passive strip CMOS detectors is fully metallised, wherefore, reflection of the laser beam is expected.
	
	The DUT is biased from the p-side (rear/back side) and readout via a n-side (front/top side) alternating current (AC) pad of a strip.
	The AC pads of the first and the second neighbouring strips are grounded, as well as the bias ring.
	The signal is amplified by a CIVIDEC~\cite{cividec} $\SI{40}{\deci\bel}$ current amplifier and recorded by an Agilent DSO9254A oscilloscope.

	\section{Analysis methods}
	\label{sec:Analysis methods}
	In the following, the later used analysis method is derived, starting from the Shockley-Ramo theorem~\cite{shockley, ramo}.
	The Shockley-Ramo theorem describes the current induced by drifting charge carriers on the collecting electrode of a sensor as:
	\begin{equation}
		\label{eqn:shockley-ramo}
		I = Q \vec{E_{\mathrm{w}}} (\mu_{\mathrm{e}} + \mu_{\mathrm{h}}) \vec{E},
	\end{equation}
	with the generated charge $Q$, the weighting field $E_{\mathrm{w}}$ that accounts for the sensor geometry, the charge carrier's mobility $\mu_{\mathrm{e/h}}$, and the electric field $E$.
	It should be noted that the charge carrier mobility depends on the electric field, the temperature, and the doping concentration~\cite{Jacoboni_Canali_Ottaviani_Alberigi_Quaranta_1977}.
	Further, the weighting field and the electric field are usually not constant across the $xy$-positions and especially along the device depth $z$, wherefore, equation~\eqref{eqn:shockley-ramo} can be noted as:
	\begin{equation}
		\label{eqn:shockley-ramo_pos}
		I(x,y,z) = Q \vec{E_{\mathrm{w}}}(x,y,z) (\mu_{\mathrm{e}} + \mu_{\mathrm{h}}) \vec{E}(x,y,z).
	\end{equation}
	The position dependence results in a varying induced current along the excess charge carriers drift, which translates to a time dependent induced current signal.
	The induced current starts with the deposition of the excess charge carriers ($t=0$) and ends as soon as they are collected ($t=t_{\mathrm{coll}}$).
	Ionising processes, like light absorption, generate excess charge carriers in pairs of electrons and holes, which have an opposite polarity and drift towards opposite collecting electrodes.
	Therefore, the measured induced current $I_{\mathrm{m}}$ is a superposition of both induced currents from electrons and holes:
	\begin{equation}
		\label{eqn:current_eh}
		I_{\mathrm{m}}(t) = I_{\mathrm{e}}(t) + I_{\mathrm{h}}(t).
	\end{equation}
	Along their drift, electrons and holes experience the weighting field and the electric field corresponding to their position and, therefore, can induce very different currents that are superimposed.
	Furthermore, their distributions are smeared out by diffusion, wherefore, the position information of the induced current can not be extracted without large uncertainties from the signal waveform.
	For this reason, equation~\eqref{eqn:shockley-ramo_pos} is investigated by using the induced current right at the deposition ($t=0$) for various depositions at different $xyz$-positions.
	However, due to a finite response of the readout electronics, the induced current at $t=0$ is experimentally not accessible and, under the assumption of a linear response of the readout electronics, the induced current at a short time after the deposition $t_{\mathrm{pc}}$ is used, to approximate $I(x,y,z)$:
	\begin{equation}
		\label{eqn:pc_approximation}
		I_{\mathrm{m}}(t_{\mathrm{pc}}, x,y,z) \approx I(x,y,z),
	\end{equation}
	where $xyz$ is the position of the excess charge carrier generation.
	This method is known as the prompt current method~\cite{kramberger_edge}.
	With respect to the investigation of detectors and especially segmented devices, it can occur that the generated excess charge carrier distribution varies, due to e.g. laser beam clipping and light reflection at metallisation or fluctuations of the laser source.
	As the prompt current method depends on the generated charge, artefacts occur if the generated charge $Q$ is not the same for all positions~\cite{PAPE2022167190}.
	To overcome the charge dependence, we propose to weight the prompt current with the collected charge, to obtain the weighted prompt current:
	\begin{equation}
		\label{eqn:wpc}
		\frac{I_{\mathrm{m}}(t_{\mathrm{pc}}, x,y,z)}{Q_{\mathrm{coll}}(x,y,z)} \approx \frac{I(x,y,z)}{Q(x,y,z)} = \vec{E_{\mathrm{w}}}(x,y,z) (\mu_{\mathrm{e}} + \mu_{\mathrm{h}}) \vec{E}(x,y,z).
	\end{equation}
	The collected charge is calculated by the integral of the measured induced current over time:
	\begin{equation}
		\label{eqn:Q}
		Q_{\mathrm{coll}}(x,y,z) = \int_0^{t_{\mathrm{coll}}}I_{\mathrm{m}}(t,x,y,z) \mathrm{d}t.
	\end{equation}
	Note that the weighted prompt current method requires that all generated charge is collected ($Q=Q_{\mathrm{coll}}$) to properly weight the prompt current.
	This condition is for example not valid if ballistic deficit is present~\cite{kolanoski2016teilchendetektoren} or if charge is lost or trapped by defects in irradiated devices~\cite{Moll_radiation}.
	
	\subsection{Extraction of the collected charge and prompt current}
\begin{figure*}
		\centering
		\begin{subfigure}{0.45\textwidth}
			\centering
			\includegraphics[width = \textwidth]{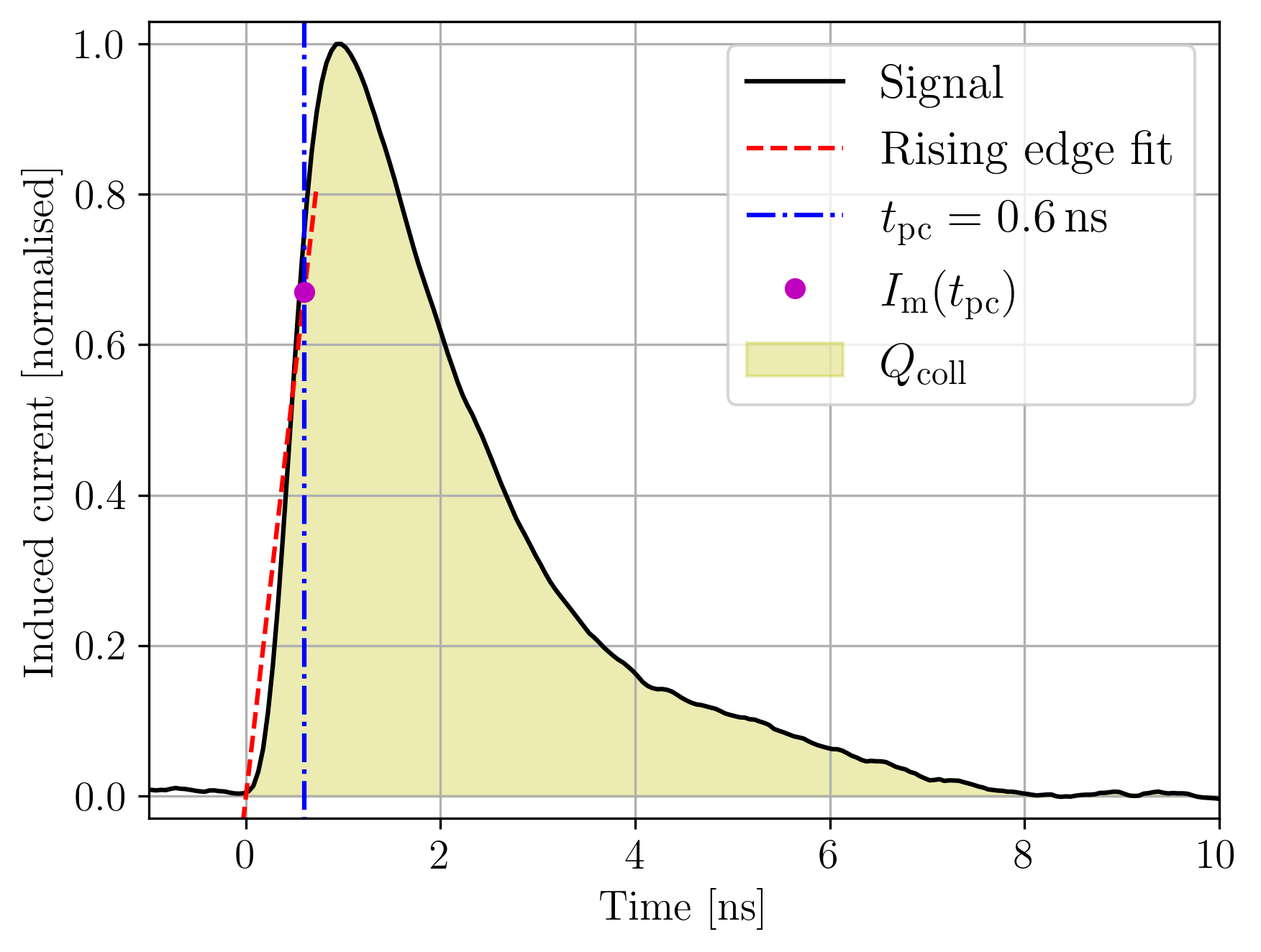}
			\caption{}	
			\label{fig:example_waveform}
		\end{subfigure}
		\begin{subfigure}{0.45\textwidth}
			\centering
			\includegraphics[width = \textwidth]{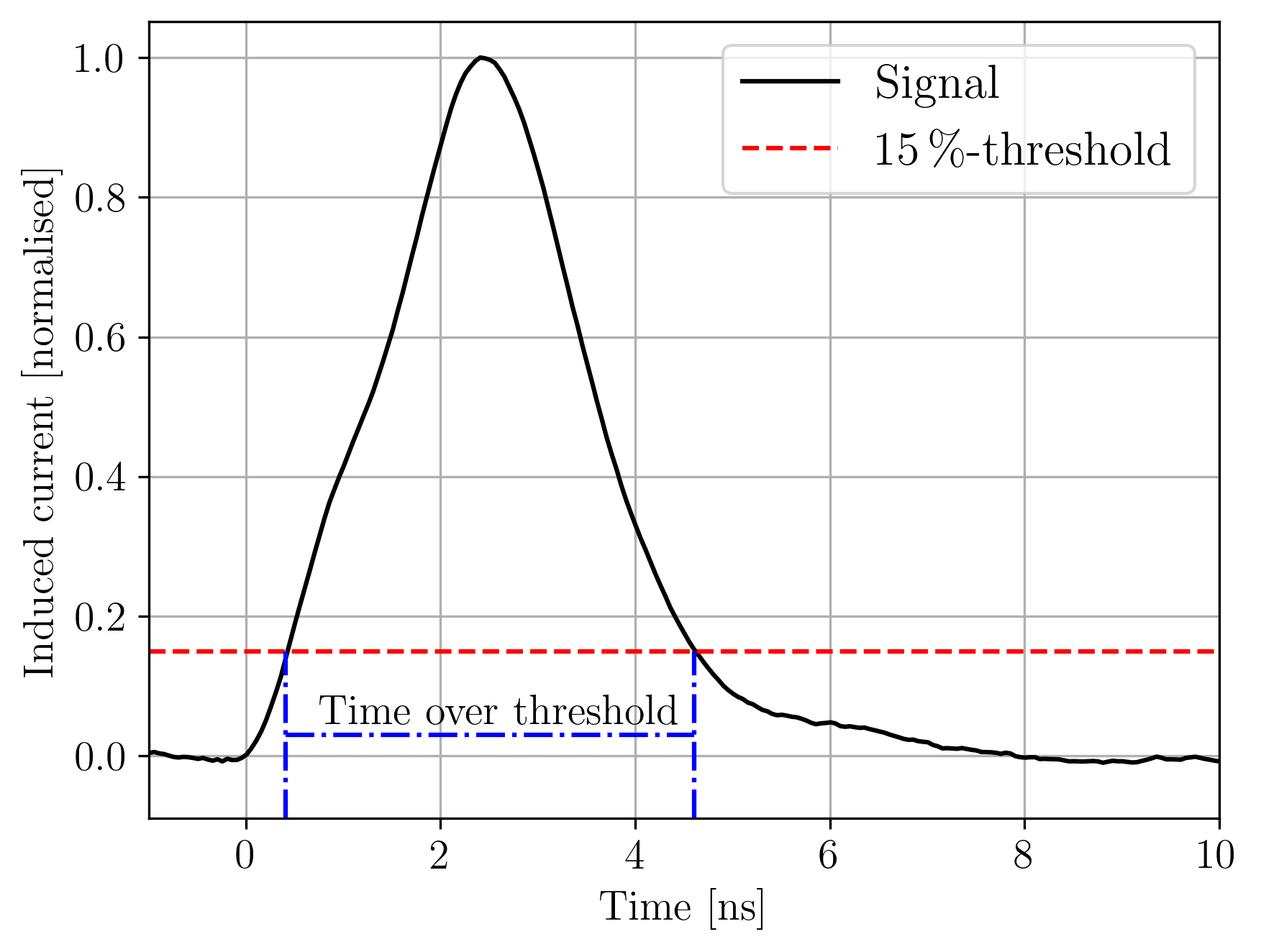}
			\caption{}
			\label{fig:tot_waveform}
		\end{subfigure}
		\caption{
			Examples of the induced current signal, with the corresponding analysis to extract the start time, the prompt current, and the collected charge~(\subref{fig:example_waveform}) and the time over threshold~(\subref{fig:tot_waveform}).
		}
		\label{fig:example_waveforms}
	\end{figure*}

	In the following, the extraction of the quantities from measurement data, needed for the weighted prompt current method, is explained. 
	The start time of an individual waveform is extracted by linear regression of the signal's rising edge.
	In a second iteration, the mean of all individual start times is taken as the start time $t=0$ that is used to extract the generated charge of the induced current, after equation~\eqref{eqn:Q}.
	A collection time of $t_{\mathrm{coll}} = \SI{10}{\nano\second}$ is used, as it is found suitable for the DUT to collect all the generated excess charge, under the used measurement conditions, i.e. bias voltage and temperature.
	Note that $t_{\mathrm{coll}}$ strongly depends on the device and the measurement conditions.
	The choice of the prompt current time $t_{\mathrm{pc}}$ is a compromise between the justification of equation~\eqref{eqn:pc_approximation} and the signal-to-noise ratio. Shorter $t_{\mathrm{pc}}$ improve the justification of the approximation, but decrease the signal-to-noise ratio.
	Following reference~\cite{kramberger_edge}, the current after $t_{\mathrm{pc}}=\SI{600}{\pico\second}$ is used as the prompt current, which is found to be a satisfactory compromise.
	Figure~\ref{fig:example_waveform} shows an example waveform, with the extraction of the prompt current and the collected charge.
	
	\subsection{Extraction of the depletion voltage and device thickness}
	Beyond the investigation of the electric field, the TPA-TCT can be used to measure the depletion voltage and the device thickness.
	Those parameters are usually extracted from a scan along the device depth for different bias voltages.
	If no laser intensity varying effects are present, the resulting charge profile $Q_{\mathrm{coll}}(z)$ can be fitted with:
	\begin{equation}
		\label{eqn:Qfit}
		Q_{\mathrm{coll}}(z) = C \left[\arctan\left(\frac{d-(z - z_{\mathrm{off}})}{z_0}\right) + \arctan\left(\frac{(z - z_{\mathrm{off}})}{z_0}\right)\right]\text{~\cite{Wiehe_paper}},
	\end{equation}
	with the thickness of the device $d$, the offset from the $z$-axis origin $z_{\mathrm{off}}$, the Rayleigh length $z_0$, and a constant $C$ that summarises material properties, setup properties, and various other constant factors.
	The device thickness is directly provided by equation~\eqref{eqn:Qfit} and the depletion voltage is found as the voltage, for which the extracted device thickness is not increasing anymore.
	The procedure for the depletion voltage is sensitive to the used collection time $t_{\mathrm{coll}}$, as the device can collect all the charge by diffusion, if it is close to full depletion and if the collection time is long enough.
	Therefore, a careful analysis of the collection time is needed to properly extract the depletion voltage with this method.
	When intensity varying effects are present, the fitting of~\eqref{eqn:Qfit} is usually not feasible along the whole device depth and the method can not be applied.
	Alternatively, the depletion voltage can be extracted from the time over threshold profile, which shows the time, the induced current signal is above a given threshold. The extraction method is shown in figure~\ref{fig:tot_waveform} for a threshold of $\SI{15}{\percent}$ of the signal's maximum.
	The threshold is calculated for each waveform individually, to avoid the influence of a varying laser intensity, because a signal with a given shape will cross a fixed threshold faster than a signal with the same shape, but smaller amplitude~\cite{leo2012techniques}. This effect is known as the time walk effect.
	In underdepleted bulk, excess carriers induce current due to the movement by diffusion, which is non-directional and, therefore, much slower than drift in depleted bulk.
	This leads to a less steep and, hence, longer signal~\cite{Kramberger_TCT}, which results in a steep rise in the time over threshold profile.
	The depletion voltage can be found as the voltage, where no influence of diffusion is present in the time over threshold profile.
	When the DUT provides reflection at the back side, the time over threshold profile can be used to extract the position of the DUT's rear side, as the profile is mirrored by this reflection~\cite{Pape_2022}.
	The closer the excess charge carriers are generated to the back side, the longer the electrons need to drift towards their collection n-electrode at the top side.
	Thus, the time over threshold increases until the reflection mirrors the profile and, hence, the position of the back side aligns with a peak in the time over threshold profile.
	The position of the front side surface is conveniently extracted from the collected charge profile by fitting~\eqref{eqn:Qfit} to only the first part of $Q(z)$.
	The device thickness is the difference between the back side and the top side position.
	
	
	\section{Results}
	\label{sec:Results}
	
	\begin{figure}
		\centering
		\includegraphics[width = 0.45\textwidth]{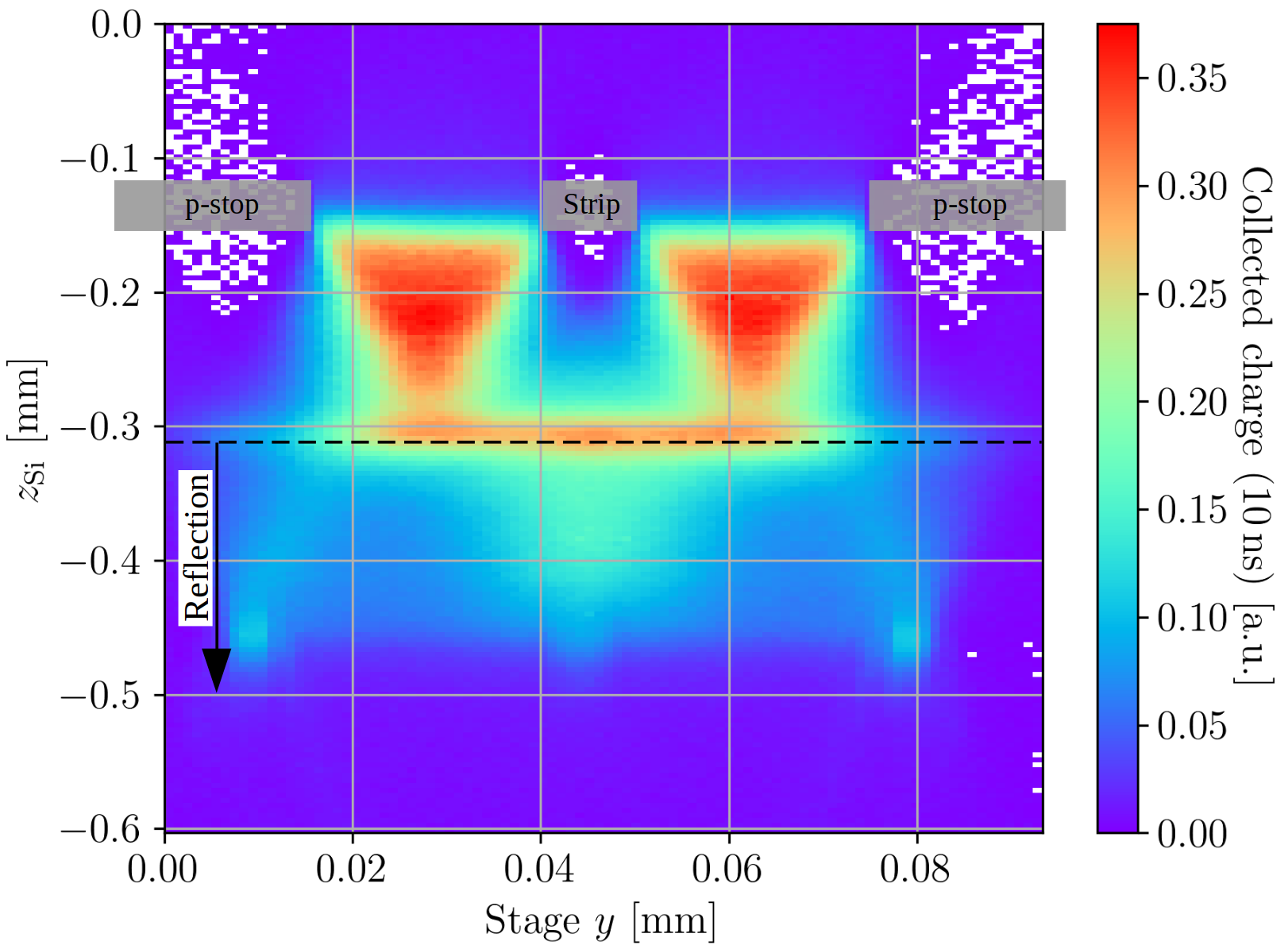}
		\caption{Charge collected within $\SI{10}{\nano\second}$ for a $yz$-scan, taken at a bias voltage of $\SI{100}{\volt}$.
			The scan is performed between the middle of the p-stop left to the readout strip ($y=\SI{0}{\milli\meter}$) and the middle of the right p-stop ($y=\SI{0.9}{\milli\meter}$).
			The top side metallisation is located at $z_{\mathrm{top}} = \SI{-0.154}{\milli\meter}$ and the backside metallisation is indicated by the black dashed line at about $z_{\mathrm{back}}=\SI{-0.31}{\milli\meter}$.
			All the signals acquired behind the back side metallisation are accounted to a focused reflection at the back side metal.
			The readout strip's metal and the metals above the p-stop are shown schematically by the transparent grey boxes.}	
		\label{fig:charge}
	\end{figure}
	In the following, the results of a $yz$-scan across the readout strip is analysed and discussed.
	The $y$-axis is oriented perpendicular to the strip metal orientation and the $z$-axis points inside the active volume.
	Figure~\ref{fig:charge} shows the collected charge for the $yz$-scan for a bias voltage of $\SI{100}{\volt}$.
	It can be seen that the profile is structured and not homogeneous below the readout strip.
	The two triangular shaped structures origin from clipping at the top side metallisations of the p-stops and the strip, and the plateau of charge collection at $z_{\mathrm{Si}}=\SI{-0.31}{\milli\meter}$ occurs due to reflection at the back side metallisation.
	It is not a feature of the device, but solely an artefact of the measurement that complicates the interpretation of the data.
	
	\begin{figure*}
		\centering
		\begin{subfigure}{0.45\textwidth}
			\centering
			\includegraphics[width = \textwidth]{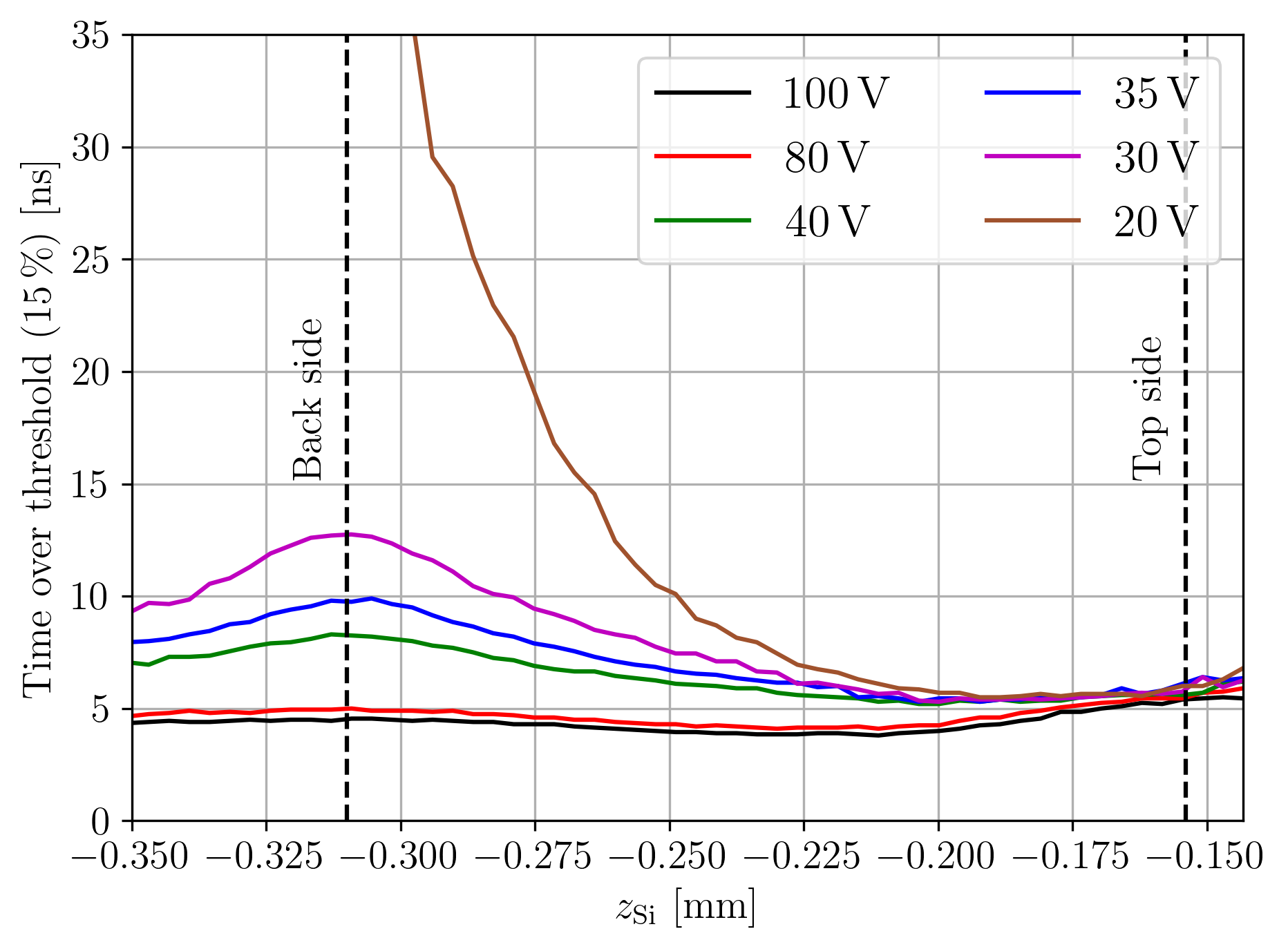}
			\caption{}
			\label{fig:tcoll15}
		\end{subfigure}
		\begin{subfigure}{0.45\textwidth}
			\centering
			\includegraphics[width = 0.99\textwidth]{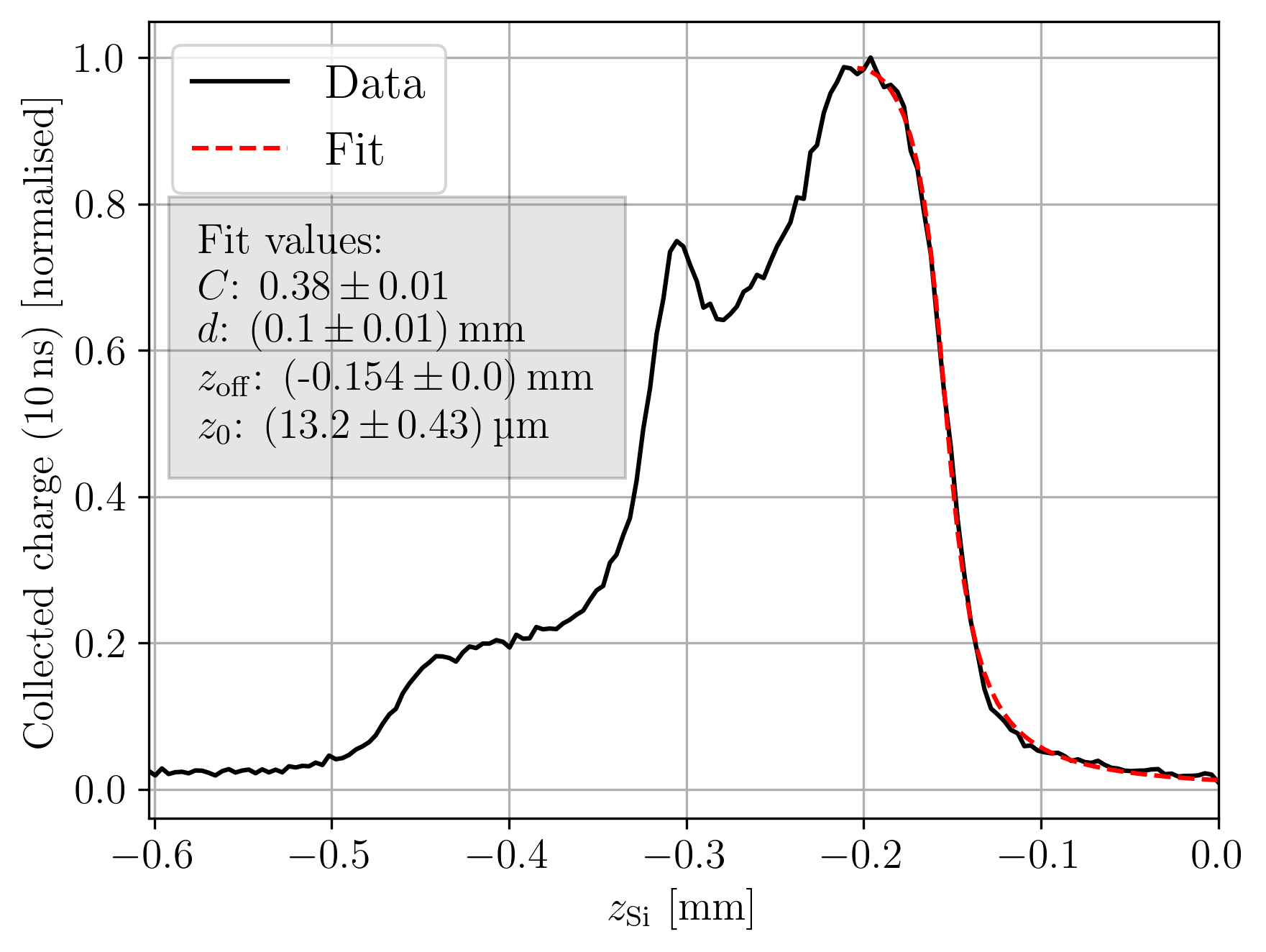}
			\caption{}
			\label{fig:qfit}
		\end{subfigure}
		\caption{
			Figure~(\subref{fig:tcoll15}) shows the time over threshold against the device depth for different bias voltages and a stage $y = \SI{0.07}{\milli\meter}$ (compare to figure~\ref{fig:charge}).
			$\SI{15}{\percent}$ of the induced current signal's maximum is used as a threshold.
			The top and back surface position are indicated by the dashed lines.
			The position of the back surface is extracted from the time over threshold profiles and the position of the top surface is extracted from a fit towards the charge collection profile shown in figure~(\subref{fig:qfit}).
			The scan is performed at the same stage $y = \SI{0.07}{\milli\meter}$ and a bias voltage of $\SI{100}{\volt}$ is used. The fit is performed according to equation~\eqref{eqn:Qfit} towards the rising edge. Note that the device thickness $d$ is not properly fitted, because only the rising edge is fitted. Due to clipping, the rest of the charge profile is not used for the fit.
		}
		\label{fig:tcoll15_qfit}
	\end{figure*}
	To extract the depletion voltage and the device positions, scans along the device depth were performed for different bias voltages at a stage $y$ of $\SI{0.07}{\milli\meter}$ (compare to figure~\ref{fig:charge}).
	The time over threshold profiles of these scans along the device depth are shown in figure~\ref{fig:tcoll15}.
	The time over threshold for $\SI{20}{\volt}$ increases steeply at $z_{\mathrm{Si}}<\SI{-0.25}{\milli\meter}$, which indicates the presents of diffusion and, thus, an incomplete depletion of the device.
	The device is fully depleted for all the other bias voltages, as no signs of diffusion are found for bias voltages $\geq \SI{30}{\volt}$.
	The position of the back side surface is extracted from the peak in the rear of the time over threshold profiles, which is found at $z_{\mathrm{back}} = \SI{-0.31}{\milli\meter}$.
	The position of the top surface is determined in figure~\ref{fig:qfit} from a fit towards the rising edge of a charge collection profile, taken after full depletion.
	Only the rising edge is fitted, because the charge collection profile is affected by clipping at the top surfaces and reflection at the back surface, which is later discussed in more detail.
	The measured Rayleigh length $z_0=\SI{13.2}{\micro\meter}$ is within the errors in agreement with the Rayleigh length measured with the knife-edge technique.
	The position of the top side is determined as $z_{\mathrm{top}} = z_{\mathrm{off}} = \SI{-0.154}{\milli\meter}$.
	Therefore, a total device thickness of $d = \SI{156}{\micro\meter}$ is found, which is in agreement with the expected thickness.
	
	\begin{figure*}
		\centering
		\begin{subfigure}{0.45\textwidth}
			\centering
			\includegraphics[width = \textwidth]{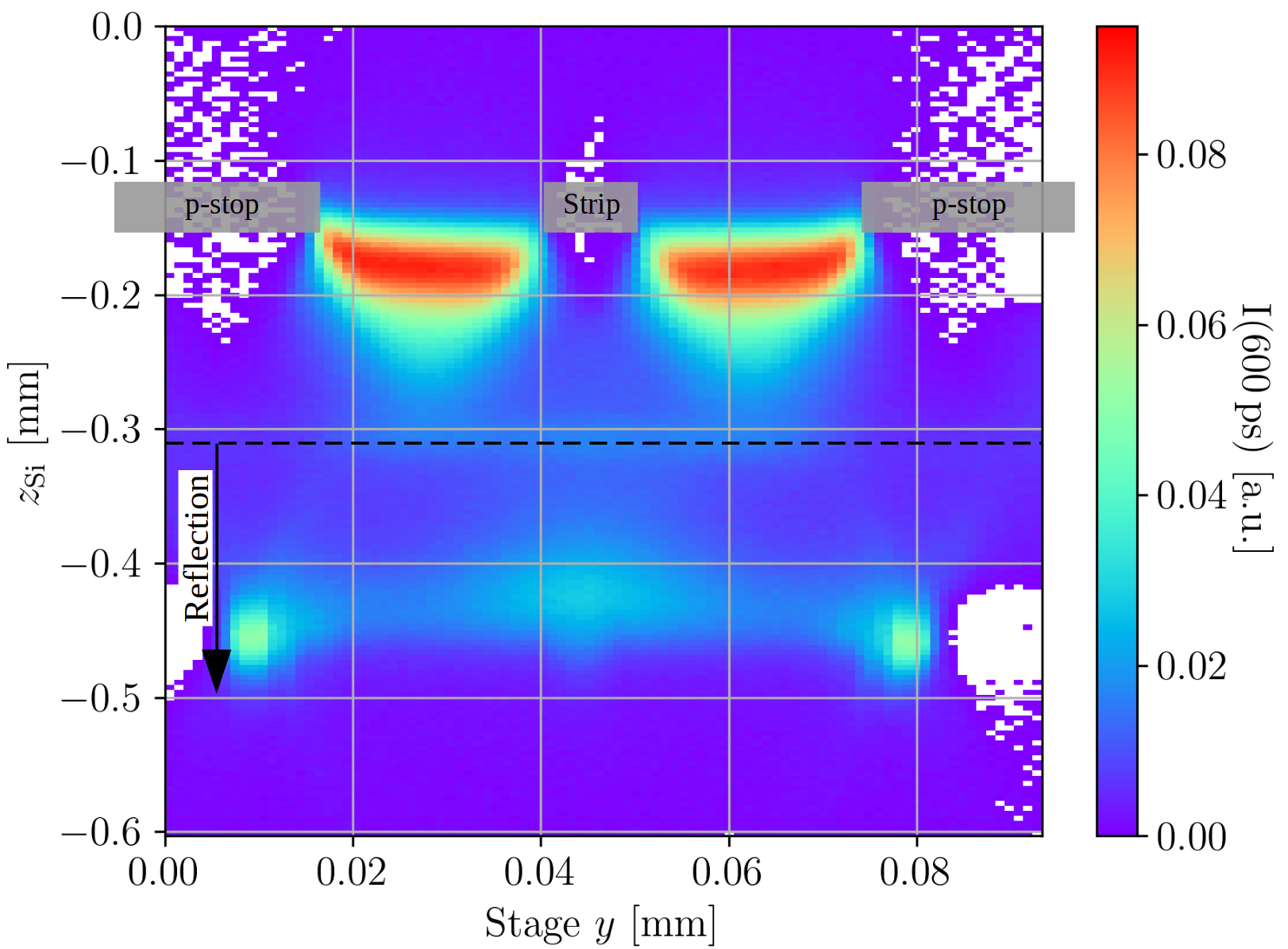}
			\caption{}
			\label{fig:pc}
		\end{subfigure}
		\begin{subfigure}{0.45\textwidth}
			\centering
			\includegraphics[width = \textwidth]{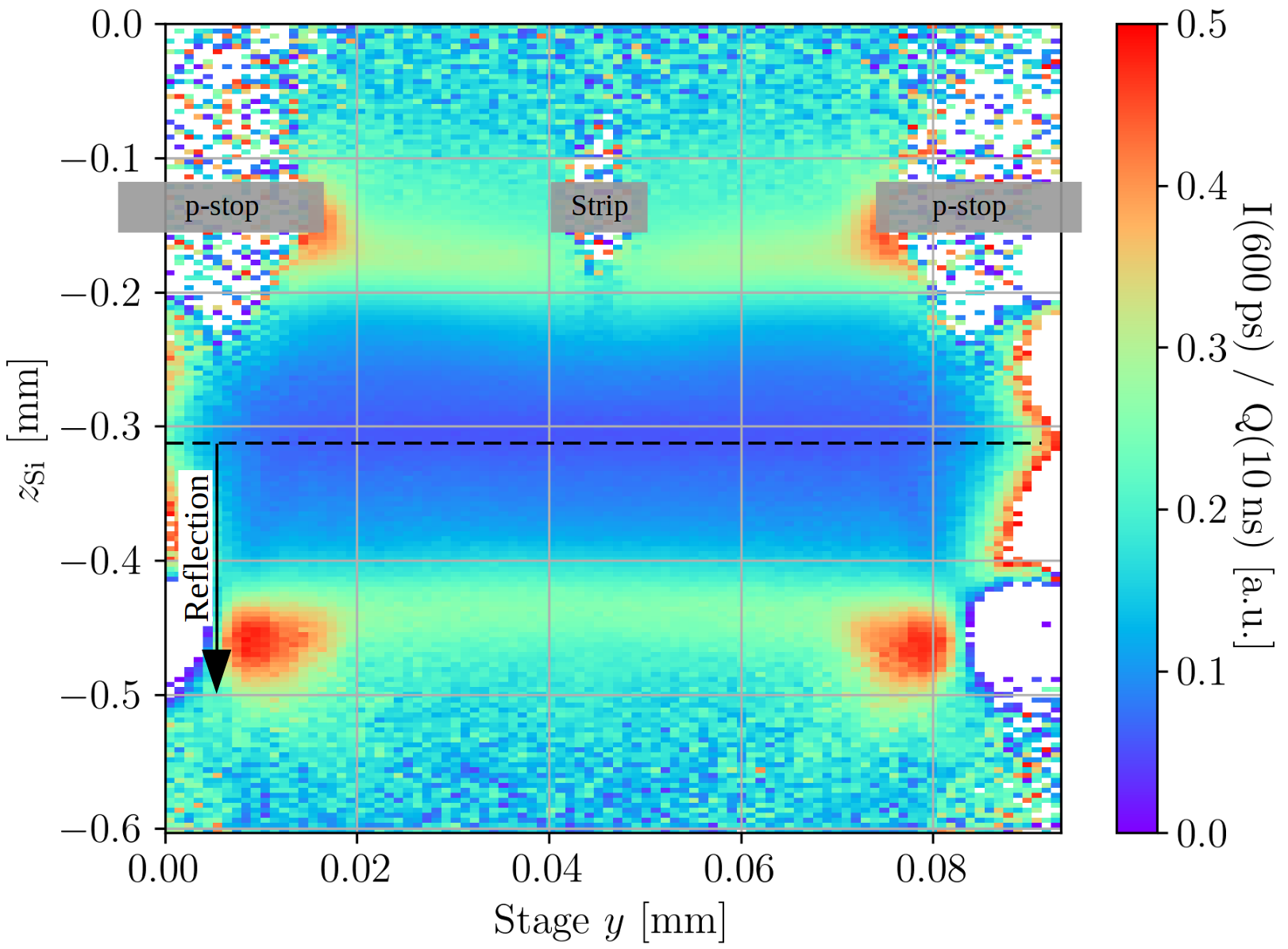}
			\caption{}
			\label{fig:wpc}
		\end{subfigure}
		\caption{
			The prompt current~(\subref{fig:pc}) and the weighted prompt current~(\subref{fig:wpc}) of the same $yz$-scan as in figure~\ref{fig:charge} are presented.
			In contrast to the prompt current and the collected charge profile, the weighted prompt current profile does not show intensity related artefacts.
			The devices back side is indicated by the black dashed line and the readout strip's metal and the metals above the p-stop are shown schematically by the transparent grey boxes.
		}
		\label{fig:pc_vs_wpc}
	\end{figure*}
\begin{figure}
		\centering
		\includegraphics[width = 0.45\textwidth]{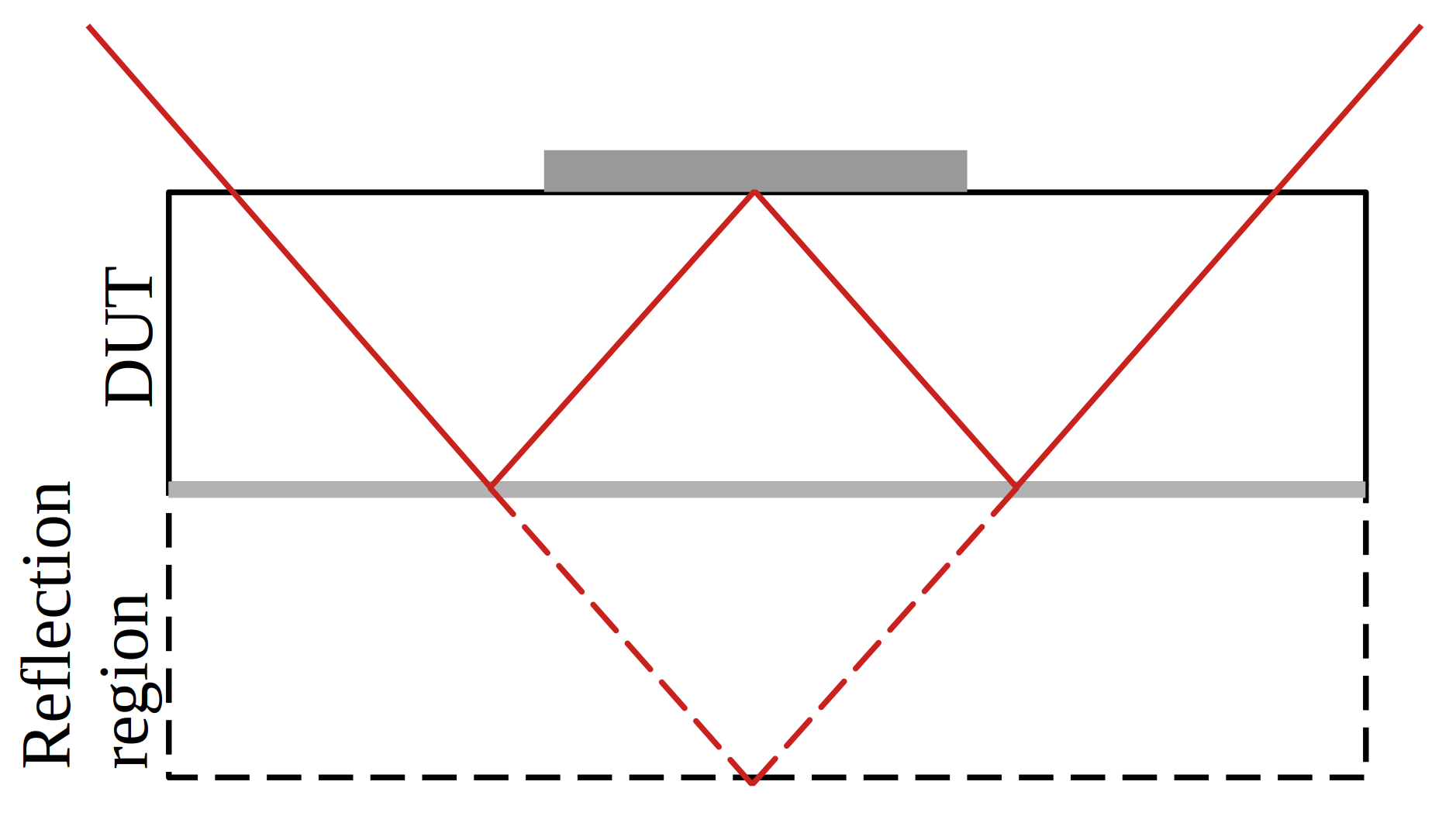}
		\caption{Schematic of the mirror technique. The reflection enables the probing below the strip metallisation with illumination from the top. The path of the outermost light rays of the laser are shown as a continuous red line. The dashed red lines show the continuation of the light rays beyond the metallised back side. Metallisations are depicted in grey.}	
		\label{fig:mirror}
	\end{figure}

	Figure~\ref{fig:pc_vs_wpc} shows a comparison between the prompt current~(\subref{fig:pc}) and the weighted prompt current~(\subref{fig:wpc}) for the same $yz$-scan of figure~\ref{fig:charge}.
	As expected, the prompt current profile contains similar structures as the collected charge profile, because the prompt current is sensitive towards changes of the collected charge.
	These artefacts are mitigated by the weighted prompt current, because it is compensated for variations of the collected charge and, therefore, not impacted by laser clipping at the top side metallisations, reflections at the back side metallisation, or any other intensity varying effect that might occur.
	The analysis is performed according to equation~\eqref{eqn:wpc} and the time $t_{\mathrm{pc}}=\SI{600}{\pico\second}$ is used.
	The observed artefacts of the charge profile fully disappear and the volume below the readout strip is much more homogeneous.
	Furthermore, it can be seen that the reflection contains meaningful information about the detector, as it is the mirror image of the DUT region.
	The reflection contains a clean picture even below the top side metallisation, wherefore, this technique can be used to investigate the electric field directly below the metal with top illumination. The mirror technique is schematically shown in figure~\ref{fig:mirror}.
	It is only available in TPA-TCT and not in SPA-TCT, because three dimensional resolution is required.
	Note that the light is always clipped by the top side metal and the clipping will vary with the position. Therefore, only light intensity independent quantities, e.g. the collection time, the weighted prompt current, or the rise time, are available to be investigated with the mirror technique.
	\section{Conclusion}
	\label{sec:Conclusion}
	In this paper, the weighted prompt current method was introduced and employed on a passive strip CMOS detector. It mitigates the intensity dependence of the prompt current method, which is useful for the investigation of segmented devices, because reflection, laser beam clipping, or any other laser intensity varying effect can be mitigated by this method. In a $yz$-scan of a passive strip CMOS detector, artefacts from clipping and reflection were observed in the collected charge profile. These artefacts fully disappear in the weighted prompt current profile, which enables the investigation of the electric field with the TPA-TCT in such devices.
	Furthermore, the mirror technique was introduced, where the reflection at the back side metallisation was exploited to probe directly below the top side metals. This technique is, depending on the metallisations, available for illumination from the top or back, but not from the edge. Besides, due to position dependent light clipping, only intensity independent quantities can be investigated with this method.
	The above introduced techniques are not only suitable for the investigation of strip detectors, but can be employed on any segmented or unsegmented detector.
	
	\section*{Acknowledgments}
	The authors would like to thank the Verbundproject 05H2021 - R\&D DETEKTOREN (Neue Trackingtechnologien): Entwicklung von aktiven und passiven mikrostrukturierten CMOS-Sensoren for providing the tested sample.
	This project was performed within the framework of RD50 and has received funding from the European Union's Horizon 2020 Research and Innovation programme under GA no 101004761 (AIDAinnova), the Wolfgang Gentner Program of the German Federal Ministry of Education and Research (grant no. 05E18CHA), and the CERN Knowledge Transfer Fund, through a grant awarded in 2017.
	
	\bibliography{references}

\end{document}